\documentclass[aps,prl,preprint,tightenlines,superscriptaddress,showpacs,byrevtex]{revtex4}
\usepackage{graphicx} 
\usepackage{dcolumn}  
\usepackage{comment}  

\graphicspath{{ps}}

\def\ulnu{\ensuremath{u \ell \nu}}
\def\btoulnu{\ensuremath{b \to u \ell \nu}}
\def\btoclnu{\ensuremath{b \to c \ell \nu}}
\def\bpiplnu{\ensuremath{B\rightarrow\pi^+\ell\nu}}
\def\bpizlnu{\ensuremath{B\rightarrow\pi^0\ell\nu}}
\def\brhoplnu{\ensuremath{B\rightarrow\rho^+\ell\nu}}
\def\brhozlnu{\ensuremath{B\rightarrow\rho^0\ell\nu}}

\def\bomegalnu{\ensuremath{B\rightarrow\omega\ell\nu}}

\def\bdplnu{\ensuremath{B\rightarrow D^{+}\ell\nu}}
\def\bdzlnu{\ensuremath{B\rightarrow D^{0}\ell\nu}}
\def\bdsplnu{\ensuremath{B\rightarrow D^{*+}\ell\nu}}
\def\bdszlnu{\ensuremath{B\rightarrow D^{*0}\ell\nu}}
\def\bxulnu{\ensuremath{B\rightarrow X_u\ell\nu}}
\def\bxclnu{\ensuremath{B\rightarrow X_c\ell\nu}}

\def\bdstarlnu{\ensuremath{B\rightarrow D^*\ell\nu}}
\def\bdlnu{\ensuremath{B\rightarrow D\ell\nu}}
\def\bddstarlnu{\ensuremath{B\rightarrow D^{**}\ell\nu}}
\def\bb{\ensuremath{B\overline{B}}\ }

\def\qsq{\ensuremath{q^2}}

\def\GeV{\ensuremath{\mathrm{GeV}}}
\def\GeVc{\ensuremath{\mathrm{GeV}/c}}
\def\GeVcc{\ensuremath{\mathrm{GeV}/c^2}}
\def\GeVGeVcc{\ensuremath{\mathrm{GeV}^2/c^2}}

\def\mm2{\ensuremath{M^2_\mathrm{miss}}}
\def\deltae{\ensuremath{\Delta E}}
\def\mbc{\ensuremath{M_\mathrm{bc}}}
\def\ebeam{\ensuremath{E_\mathrm{beam}}}
\def\pb{\ensuremath{\vec{p}_B}}
\def\btag{\ensuremath{B_\mathrm{tag}}}

\def\pipresultqsqafit{  \ensuremath{0.50 \pm 0.14}}
\def\pizresultqsqafit{  \ensuremath{0.28 \pm 0.09}}

\def\pipresultqsqbfit{  \ensuremath{0.68 \pm 0.18}}
\def\pizresultqsqbfit{  \ensuremath{0.22 \pm 0.08}}

\def\pipresultqsqcfit{  \ensuremath{0.31 \pm 0.12}}
\def\pizresultqsqcfit{  \ensuremath{0.36 \pm 0.12}}

\def\pipresultqsqsumfit{  \ensuremath{1.49 \pm 0.26}}
\def\pizresultqsqsumfit{  \ensuremath{0.86 \pm 0.17}}

\def\pipsyserrqsqafit{  \ensuremath{0.02}}
\def\pizsyserrqsqafit{  \ensuremath{0.02}}

\def\pipsyserrqsqbfit{  \ensuremath{0.03}}
\def\pizsyserrqsqbfit{  \ensuremath{0.04}}

\def\pipsyserrqsqcfit{  \ensuremath{0.01}}
\def\pizsyserrqsqcfit{  \ensuremath{0.02}}

\def\pipsyserrqsqsumfit{  \ensuremath{0.06}}
\def\pizsyserrqsqsumfit{  \ensuremath{0.06}}

\def\pipshortresultqsqafit{  \ensuremath{\pipresultqsqafit\ \pm
                                         \pipsyserrqsqafit}}
\def\pizshortresultqsqafit{  \ensuremath{\pizresultqsqafit\ \pm
                                         \pizsyserrqsqafit}}

\def\pipshortresultqsqbfit{  \ensuremath{\pipresultqsqbfit\ \pm
                                         \pipsyserrqsqbfit}}
\def\pizshortresultqsqbfit{  \ensuremath{\pizresultqsqbfit\ \pm
                                         \pizsyserrqsqbfit}}

\def\pipshortresultqsqcfit{  \ensuremath{\pipresultqsqcfit\ \pm
                                         \pipsyserrqsqcfit}}
\def\pizshortresultqsqcfit{  \ensuremath{\pizresultqsqcfit\ \pm
                                         \pizsyserrqsqcfit}}

\def\pipshortresultqsqsumfit{  \ensuremath{\pipresultqsqsumfit\ \pm
                                           \pipsyserrqsqsumfit}}
\def\pizshortresultqsqsumfit{  \ensuremath{\pizresultqsqsumfit\ \pm
                                           \pizsyserrqsqsumfit}}

\begin{document}


\preprint{\vbox{ \hbox{   }
                 \hbox{BELLE-CONF-0666}
                 \hbox{16th October 2006}
}}

\title{ \quad\\[0.5cm]  Measurement of exclusive $B\to X_u \ell \nu$
  decays using a full-reconstruction tag at Belle }


\affiliation{Budker Institute of Nuclear Physics, Novosibirsk}
\affiliation{Chiba University, Chiba}
\affiliation{Chonnam National University, Kwangju}
\affiliation{University of Cincinnati, Cincinnati, Ohio 45221}
\affiliation{University of Frankfurt, Frankfurt}
\affiliation{The Graduate University for Advanced Studies, Hayama} 
\affiliation{Gyeongsang National University, Chinju}
\affiliation{University of Hawaii, Honolulu, Hawaii 96822}
\affiliation{High Energy Accelerator Research Organization (KEK), Tsukuba}
\affiliation{Hiroshima Institute of Technology, Hiroshima}
\affiliation{University of Illinois at Urbana-Champaign, Urbana, Illinois 61801}
\affiliation{Institute of High Energy Physics, Chinese Academy of Sciences, Beijing}
\affiliation{Institute of High Energy Physics, Vienna}
\affiliation{Institute of High Energy Physics, Protvino}
\affiliation{Institute for Theoretical and Experimental Physics, Moscow}
\affiliation{J. Stefan Institute, Ljubljana}
\affiliation{Kanagawa University, Yokohama}
\affiliation{Korea University, Seoul}
\affiliation{Kyoto University, Kyoto}
\affiliation{Kyungpook National University, Taegu}
\affiliation{Swiss Federal Institute of Technology of Lausanne, EPFL, Lausanne}
\affiliation{University of Ljubljana, Ljubljana}
\affiliation{University of Maribor, Maribor}
\affiliation{University of Melbourne, Victoria}
\affiliation{Nagoya University, Nagoya}
\affiliation{Nara Women's University, Nara}
\affiliation{National Central University, Chung-li}
\affiliation{National United University, Miao Li}
\affiliation{Department of Physics, National Taiwan University, Taipei}
\affiliation{H. Niewodniczanski Institute of Nuclear Physics, Krakow}
\affiliation{Nippon Dental University, Niigata}
\affiliation{Niigata University, Niigata}
\affiliation{University of Nova Gorica, Nova Gorica}
\affiliation{Osaka City University, Osaka}
\affiliation{Osaka University, Osaka}
\affiliation{Panjab University, Chandigarh}
\affiliation{Peking University, Beijing}
\affiliation{University of Pittsburgh, Pittsburgh, Pennsylvania 15260}
\affiliation{Princeton University, Princeton, New Jersey 08544}
\affiliation{RIKEN BNL Research Center, Upton, New York 11973}
\affiliation{Saga University, Saga}
\affiliation{University of Science and Technology of China, Hefei}
\affiliation{Seoul National University, Seoul}
\affiliation{Shinshu University, Nagano}
\affiliation{Sungkyunkwan University, Suwon}
\affiliation{University of Sydney, Sydney NSW}
\affiliation{Tata Institute of Fundamental Research, Bombay}
\affiliation{Toho University, Funabashi}
\affiliation{Tohoku Gakuin University, Tagajo}
\affiliation{Tohoku University, Sendai}
\affiliation{Department of Physics, University of Tokyo, Tokyo}
\affiliation{Tokyo Institute of Technology, Tokyo}
\affiliation{Tokyo Metropolitan University, Tokyo}
\affiliation{Tokyo University of Agriculture and Technology, Tokyo}
\affiliation{Toyama National College of Maritime Technology, Toyama}
\affiliation{University of Tsukuba, Tsukuba}
\affiliation{Virginia Polytechnic Institute and State University, Blacksburg, Virginia 24061}
\affiliation{Yonsei University, Seoul}
  \author{K.~Abe}\affiliation{High Energy Accelerator Research Organization (KEK), Tsukuba} 
  \author{K.~Abe}\affiliation{Tohoku Gakuin University, Tagajo} 
  \author{I.~Adachi}\affiliation{High Energy Accelerator Research Organization (KEK), Tsukuba} 
  \author{H.~Aihara}\affiliation{Department of Physics, University of Tokyo, Tokyo} 
  \author{D.~Anipko}\affiliation{Budker Institute of Nuclear Physics, Novosibirsk} 
  \author{K.~Aoki}\affiliation{Nagoya University, Nagoya} 
  \author{T.~Arakawa}\affiliation{Niigata University, Niigata} 
  \author{K.~Arinstein}\affiliation{Budker Institute of Nuclear Physics, Novosibirsk} 
  \author{Y.~Asano}\affiliation{University of Tsukuba, Tsukuba} 
  \author{T.~Aso}\affiliation{Toyama National College of Maritime Technology, Toyama} 
% ...
\collaboration{The Belle Collaboration}

\noaffiliation

\begin{abstract}

  We report a preliminary study of the branching fractions and $q^2$
  distributions of exclusive charmless semileptonic $B$ decays,
  using events tagged by fully reconstructing one of
  the $B$ mesons in a hadronic decay mode. 
  These results are obtained from a data sample that contains 535
  $\times 10^6\ B\bar{B}$ pairs, collected 
  near the $\Upsilon(4S)$ resonance
  with the Belle detector at the KEKB asymmetric energy $e^+ e^-$
  collider.
\end{abstract}


\maketitle

\tighten

{\renewcommand{\thefootnote}{\fnsymbol{footnote}}}
\setcounter{footnote}{0}


The Standard Model (SM) of particle physics contains a number of
parameters whose values are not predicted by theory and must therefore
be measured by experiment. In the quark sector, the elements of the
Cabibbo-Kobayashi-Maskawa (CKM) matrix govern
the weak transitions between quark flavours, and precision
measurements of their values are desirable. In particular, much
experimental and theoretical effort is currently being employed to
test the consistency of the CKM formalism \cite{KM},  by examining
the Unitarity Triangle most relevant to the decays of $B$ mesons.

The precision to which the angle
$\sin 2\phi_1$ characterising indirect $CP$ violation in 
$b \to c \overline{c} s$ transitions has improved to approximately
5\% \cite{hara_fpcp06}. This makes a precision measurement of the
length of the side of
the Unitarity triangle opposite to $\phi_1$ particularly
important as a consistency check of the SM picture. The length of this
side is determined to good approximation by the ratio of the
magnitudes of two CKM matrix elements, $|V_{ub}|/|V_{cb}|$. Both of
these can be measured using exclusive semileptonic $B$ meson decays.
Using charmed semileptonic decays, the precision to which $|V_{cb}|$
has been determined is of order 2\%. On the other hand $|V_{ub}|$,
which can be measured using charmless semileptonic decays, is the most
poorly known of the CKM matrix elements. Both inclusive and exclusive
methods of measuring $|V_{ub}|$ have been pursued, with the inclusive
methods giving a value to a precision of 7-8\%. The exclusive
determination of $|V_{ub}|$ currently has a precision poorer than
10\%. It is the aim of an ongoing programme of measurements at the
$B$ factories to improve this precision to better than 5\%, for
comparison with the inclusive results, which have somewhat different
experimental and theoretical systematics. This would provide a sharp
consistency test with the value of $\sin 2\phi_1$.

In this paper we present preliminary studies of the exclusive
semileptonic decays \bpiplnu, \bpizlnu,
\brhoplnu, \brhozlnu\ and \bomegalnu\ using a full reconstruction
tagging technique to identify candidate $B$ mesons. 
This measurement
is based on a data sample that
contains 535 $\times 10^6 B\overline{B}$ pairs, 
collected  with the Belle detector at the KEKB asymmetric-energy
$e^+e^-$ (3.5 on 8~GeV) collider~\cite{KEKB}.
KEKB operates at the $\Upsilon(4S)$ resonance 
($\sqrt{s}=10.58$~GeV) with a peak luminosity that exceeds
$1.6\times 10^{34}~{\rm cm}^{-2}{\rm s}^{-1}$.
The $\Upsilon(4S)$ is produced
with a Lorentz boost of $\beta\gamma=0.425$ nearly along
the electron beamline ($z$).


The Belle detector is a large-solid-angle magnetic
spectrometer that
consists of a silicon vertex detector (SVD),
a 50-layer central drift chamber (CDC), an array of
aerogel threshold \v{C}erenkov counters (ACC), 
a barrel-like arrangement of time-of-flight
scintillation counters (TOF), and an electromagnetic calorimeter
comprised of CsI(Tl) crystals (ECL) located inside 
a super-conducting solenoid coil that provides a 1.5~T
magnetic field.  An iron flux-return located outside of
the coil is instrumented to detect $K_L^0$ mesons and to identify
muons (KLM).  The detector
is described in detail elsewhere~\cite{Belle}.
Two inner detector configurations were used. A 2.0 cm beampipe
and a 3-layer silicon vertex detector was used for the first sample
of 152 $\times 10^6\ B\bar{B}$ pairs, while a 1.5 cm beampipe, a 4-layer
silicon detector and a small-cell inner drift chamber were used to record  
the remaining 383 $\times 10^6 B\bar{B}$ pairs\cite{svd}.  


Table~\ref{tab:b_to_ulnu} lists all current measurements of
branching fractions for exclusive \bxulnu\ decays, where $X_u$ denotes
a light meson containing a $u$ quark. This table is based
on a recent compilation by the Heavy Flavour Averaging Group (HFAG)
\cite{HFAG}. Three methods of identifying signal candidates
have been employed in these measurements, denoted ``untagged'',
``semileptonic tagged'' or ``full reconstruction tagged''. Only BaBar
have reported results to date based on the third of these
\cite{babar-fullrecon-pi-05} \cite{babar-fullrecon-04}.

The most precise measurements at the present time come from
the untagged analyses of CLEO \cite{cleo-untagged-03} and BaBar
\cite{babar-untagged-05}. As more integrated luminosity is accumulated
by the $B$-factory experiments, the full reconstruction tagging
technique will become the most precise method. It holds the advantage
of providing the best signal to background ratio, offset by the lowest
efficiencies.

\begin{table}[htb]
\caption{Measurements of branching fractions of
  exclusive $B \to X_u \ell \nu$ decay modes.
    In each case, the first error is
    statistical, the second experimental systematic, the third due to
    form factor uncertainties for the signal mode, and the fourth, when
    present, due to form factor uncertainties for crossfeed modes.
    U indicates untagged method, S semileptonic tagging method, and F
    full reconstruction tagging.}
\label{tab:b_to_ulnu} 
\begin{tabular}
      {@{\hspace{0.5cm}}l@{\hspace{0.5cm}} ||
       @{\hspace{0.5cm}}c@{\hspace{0.5cm}} ||
       @{\hspace{0.5cm}}c@{\hspace{0.5cm}} ||
       @{\hspace{0.5cm}}c@{\hspace{0.5cm}}}
\hline \hline
  \textbf{Experiment} & \textbf{Mode} &
  \textbf{$B\overline{B}$} &
\textbf{Branching Fraction} \\
 \textbf{and Tag Method} & &  
\textbf{$\left[ 10^{6} \right]$} &
\textbf{$\left[ 10^{-4} \right]$}
\\
\hline
CLEO \cite{cleo-untagged-03} \hfill U &  $B^0 \to \pi^- \ell \nu$ &
9.7 & 
$1.33 \pm 0.18 \pm 0.11 \pm 0.01 \pm 0.07$ \\
BaBar \cite{babar-untagged-05} \hfill U & $B^0 \to \pi^- \ell \nu$ &
86 &
$1.38 \pm 0.10 \pm 0.16
      \pm 0.08$ \\
Belle \cite{belle-sl-06} \hfill S & $B^0 \to \pi^- \ell \nu$ &
275 & 
$1.38 \pm 0.19 \pm 0.14
      \pm 0.03$ \\
Belle \cite{belle-sl-06} \hfill S  & $B^+ \to \pi^0 \ell \nu$ &
275 & 
$0.77 \pm 0.14 \pm 0.08
      \pm 0.00$ \\
BaBar \cite{babar-sl-pip-05} \hfill S  & $B^0 \to \pi^- \ell \nu$ &
232 & 
$1.03 \pm 0.25 \pm 0.13$ \\
BaBar \cite{babar-sl-piz-05} \hfill S  & $B^+ \to \pi^0 \ell \nu$ &
88 & 
$1.80 \pm 0.37 \pm 0.23$ \\
BaBar \cite{babar-fullrecon-pi-05} \hfill F & $B^0 \to \pi^- \ell \nu$ &
233 & 
$1.14 \pm 0.27 \pm 0.17$ \\
BaBar \cite{babar-fullrecon-pi-05} \hfill F & $B^+ \to \pi^0 \ell \nu$ &
233 & 
$0.86 \pm 0.22 \pm 0.11$ \\
BaBar \cite{babar-fullrecon-pi-05} \hfill F & $B \to \pi \ell \nu$ &
233 & 
$1.28 \pm 0.23 \pm 0.16$ \\
CLEO \cite{cleo-untagged-03} \hfill U &  $B^0 \to \rho^- \ell \nu$ &
9.7 &
$2.17 \pm 0.34 ^{+0.47}_{-0.54}
\pm 0.41 \pm 0.01$ \\
CLEO \cite{cleo-untagged-00} \hfill U &  $B^0 \to \rho^- \ell \nu$ &
3.3 &
$2.69 \pm 0.41 ^{+0.35}_{-0.40}
\pm 0.50$ \\
BaBar \cite{babar-fullrecon-04} \hfill F & $B^0 \to \rho^- \ell \nu$ &
88 &
$2.57 \pm 0.52 \pm 0.59$ \\
BaBar \cite{babar-untagged-03} \hfill U & $B^0 \to \rho^- e \nu$ &
55 &
$3.29 \pm 0.42 \pm 0.47
\pm 0.60$ \\
BaBar \cite{babar-untagged-05} \hfill U & $B^0 \to \rho^- \ell \nu$ &
83 &
$2.14 \pm 0.21 \pm 0.51
\pm 0.28$ \\
Belle \cite{belle-sl-06} \hfill S & $B^0 \to \rho^- \ell \nu$ &
275 &
$2.17 \pm 0.54 \pm 0.31
\pm 0.08$ \\
Belle \cite{belle-sl-06} \hfill S & $B^+ \to \rho^0 \ell \nu$ &
275 & 
$1.33 \pm 0.23 \pm 0.17
\pm 0.05$ \\
Belle \cite{belle-untagged-omega-04} \hfill U & $B^+ \to \omega \ell \nu$ &
85 &
$1.3 \pm 0.4 \pm 0.2
\pm 0.3$ \\
CLEO \cite{cleo-untagged-03} \hfill U &  $B^+ \to \eta \ell \nu$ &
9.7 &
$0.84 \pm 0.31 \pm 0.16
\pm 0.09$ \\
\hline \hline
\end{tabular}
\end{table}


In this analysis we fully reconstruct one of the two $B$ mesons from
the $\Upsilon(4S)$ decay
(\btag) in one of the following hadronic decay modes,
$B^- \to D^{(*)0} \pi^-$, $B^- \to D^{(*)0} \rho^-$, $B^- \to D^{(*)0} a_1^-$,
$B^- \to D^{(*)0} D_s^{(*)-}$,
$B^0 \to D^{(*)+} \pi^-$, $B^0 \to D^{(*)+} \rho^-$, $B^0 \to D^{(*)+} a_1^-$
or
$B^0 \to D^{(*)+} D_s^{(*)-}$ \cite{CC}. Decays are identified on the basis of
the proximity of the beam-energy constrained mass
$\mbc = \sqrt{(\ebeam)^2 - (\pb)^2}$ and $\deltae = E_B - \ebeam$ to
their nominal values of the $B$ meson rest mass and zero,
respectively.
Here \ebeam, \pb\ and $E_B$ are the beam energy and the measured
3-momentum and energy of the \btag\ candidate in the $\Upsilon(4S)$
rest frame respectively. If multiple tag candidates are found, the one with
values of \mbc\ and \deltae\ closest to nominal is chosen. Events with
a \btag\ satisfying the selections $\mbc > 5.27\ \GeVcc$ and
$-0.08 < \deltae < 0.06\ \GeV\ $ are retained. The charge of the \btag\
candidate is necessarily restricted to $Q_\mathrm{tag} = 0$
or $Q_\mathrm{tag} = \pm 1$
by demanding that it is consistent with one of the above decay modes.


Reconstructed charged tracks and ECL clusters which are not associated with the
\btag\ candidate are used to search for the signal $B$ meson decays of
interest recoiling against the \btag. Photons identified with isolated
ECL clusters which have a laboratory energy of less than 50 MeV are ignored.

Electrons are identified using information on $dE/dx$ from the CDC,
response of the ACC, shower shape in the ECL, and the ratio of the
energy deposited in the ECL to the momentum determined from tracking.
The signals in the KLM are used to identify muons. Charged kaons are
identified based on the $dE/dx$ information from the CDC, the
\v{C}erenkov light yields in the ACC and time-of-flight information from
the TOF counters. Any charged particles which are not identified as
leptons or kaons are taken to be pions.

Photons whose direction in the laboratory frame lies within a $5^\circ$ cone 
of the direction of an identified lepton are considered to be
bremsstrahlung. The 4-momentum of the photon is added to that of the
lepton and the photon is not considered further.

Neutral pions are reconstructed from pairs of photons whose invariant
mass lies in the range $[0.120, 0.150]$\ \GeVcc, of order $\pm 3 \sigma$
of the $\pi^0$ mass. Charged  $\rho$ meson
candidates are reconstructed via the decay $\rho^\pm \to \pi^\pm$
$\pi^0$ where the invariant mass of the pair of pions is required to
lie in the range $[0.570, 0.970]$\ \GeVcc. Neutral $\rho$ meson candidates
are similarly reconstructed from pairs of oppositely charged pions,
with the requirement that $m_{\pi^+ \pi^-}$ is in the range
$[0.703, 0.863]$\ \GeVcc. Finally, $\omega$ candidates are
reconstructed from $\omega \to \pi^+$ $\pi^-$ $\pi^0$ with
$m_{\pi^+ \pi^- \pi^0}$ in the range $[0.690, 0.850]$\ \GeVcc.
In events where more than one hadron candidate, denoted $X_u$, of a
given type is identified amongst the recoil particles, the candidate
with the highest momentum in the $\Upsilon(4S)$ rest frame is chosen.

To isolate signal candidates, several requirements are placed on
the recoil system. There must be one lepton candidate present only,
with $p_{lep} > 0.4\ \GeVc$ in the $\Upsilon(4S)$ rest frame. The
total charge of the recoil system, $Q_\mathrm{recoil}$, is required to
be $0$ if a
neutral tag has been identified, and $\pm 1$ if a charged tag has been
identified. In the charged case, the sign of $Q_\mathrm{recoil}$ must be
opposite to that of $Q_\mathrm{tag}$. In the neutral case, we do not make any
requirement on the sign of the lepton charge with respect to the
\btag, to allow for mixing.

The number of charged recoil particles
is required to correspond to one of the sought modes, i.e. one for
\bpizlnu\ (the lepton), two for \bpiplnu\ and \brhoplnu\ (the lepton
plus one charged pion) and three for \brhozlnu\ and \bomegalnu\ (the
lepton plus two charged pions). Additionally, the number of recoil
$\pi^0$ candidates is required to be consistent with one of the sought
modes. In order to increase efficiency, however, we allow more than
the necessary number in some cases:  we require
no $\pi^0$ candidates to be present for \bpiplnu\ and \brhozlnu\ modes,
and at least one $\pi^0$ for the \bpizlnu, \brhoplnu\ and \bomegalnu\
modes. Additionally, we require that there be no more than $0.3\ \GeV\ $
of residual neutral energy present on the recoil side in the
$\Upsilon(4S)$ rest frame, after any photons
contributing to the $X_u$ candidate have been removed.


Signal events are identified by examining the missing mass squared
(\mm2) distributions. If the tagging $B$ is correctly
reconstructed and the correct lepton and hadron candidate have been
identified on the recoil side, then (ideally) all missing
4-momentum is due to the remaining unidentified neutrino.
The square of the missing 4-momentum for signal events should
therefore be close to zero, and applying this requirement provides a
very strong discrimination between signal and background.


Background contributions come from several sources. These include
semileptonic decays resulting from \btoclnu\ transitions, denoted
\bxclnu, which have significantly larger branching fractions to the
channels under study; continuum $e^+ + e^- \to q\overline{q}$
processes; and cross feed from one \bxulnu\ channel into another.
The contributions of these backgrounds are studied
using Monte Carlo (MC) simulated data samples generated with the
EvtGeb package \cite{evtgen}.
Generic \bb and continuum MC samples equivalent to
approximately twice the integrated luminosity of the real data set
are used. The model
adopted for \bdstarlnu\ and \bdlnu\ decays is based on HQET and
parametrisation of the form factors \cite{caprini}, while \bddstarlnu\
decays are based on the ISGW2 model \cite{isgw2}. A non-resonant
$B \to D^{(*)}\pi \ell \nu$ component based on the Goity-Roberts prescription
\cite{goity} is also included.

A separate MC sample equivalent to approximately seven times the integrated
luminosity of the real data set is used to simulate the signal
channels and crossfeed from other \bxulnu\ decays. Models for the
exclusive modes are based on Light Cone Sum Rules (LSCR) for
$\pi$ \cite{ball-and-zwicky-01}, $\rho$ and $\omega$
\cite{ball-and-braun-98} modes and ISGW2 \cite{isgw2}
for other exclusive modes.

Radiative effects associated with the lepton and resulting from
higher-order QED processes are modelled in all MC
samples using the PHOTOS package \cite{photos}. All generated MC
events are passed through a full simulation of Belle detector effects
based on GEANT 3.21 \cite{geant}.

\begin{figure}[h]
\label{fig:mm2fit}
\begin{tabular}{ll}
  (a) & (b) \\
\includegraphics[width=0.35\textwidth]{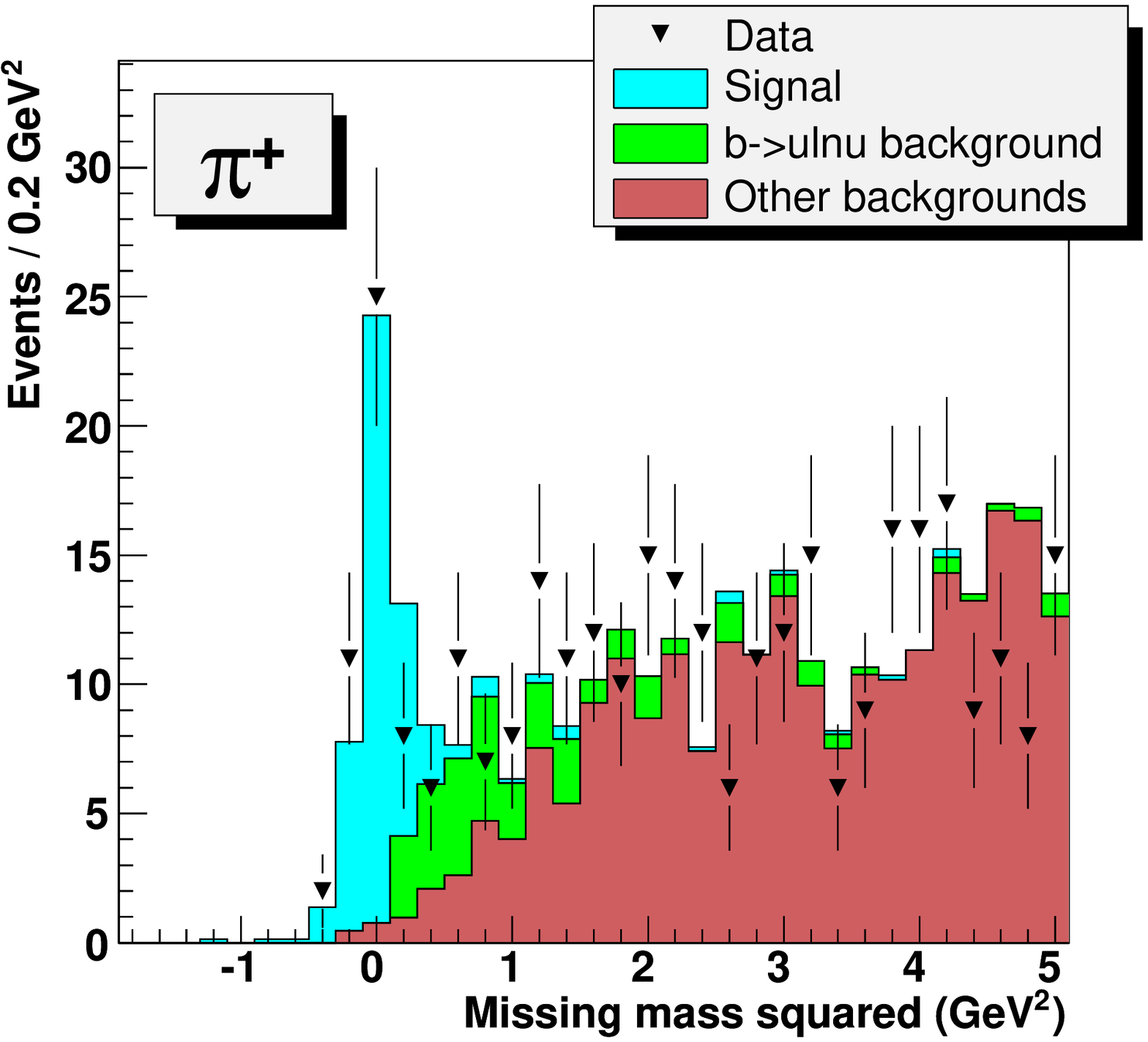} &
\includegraphics[width=0.35\textwidth]{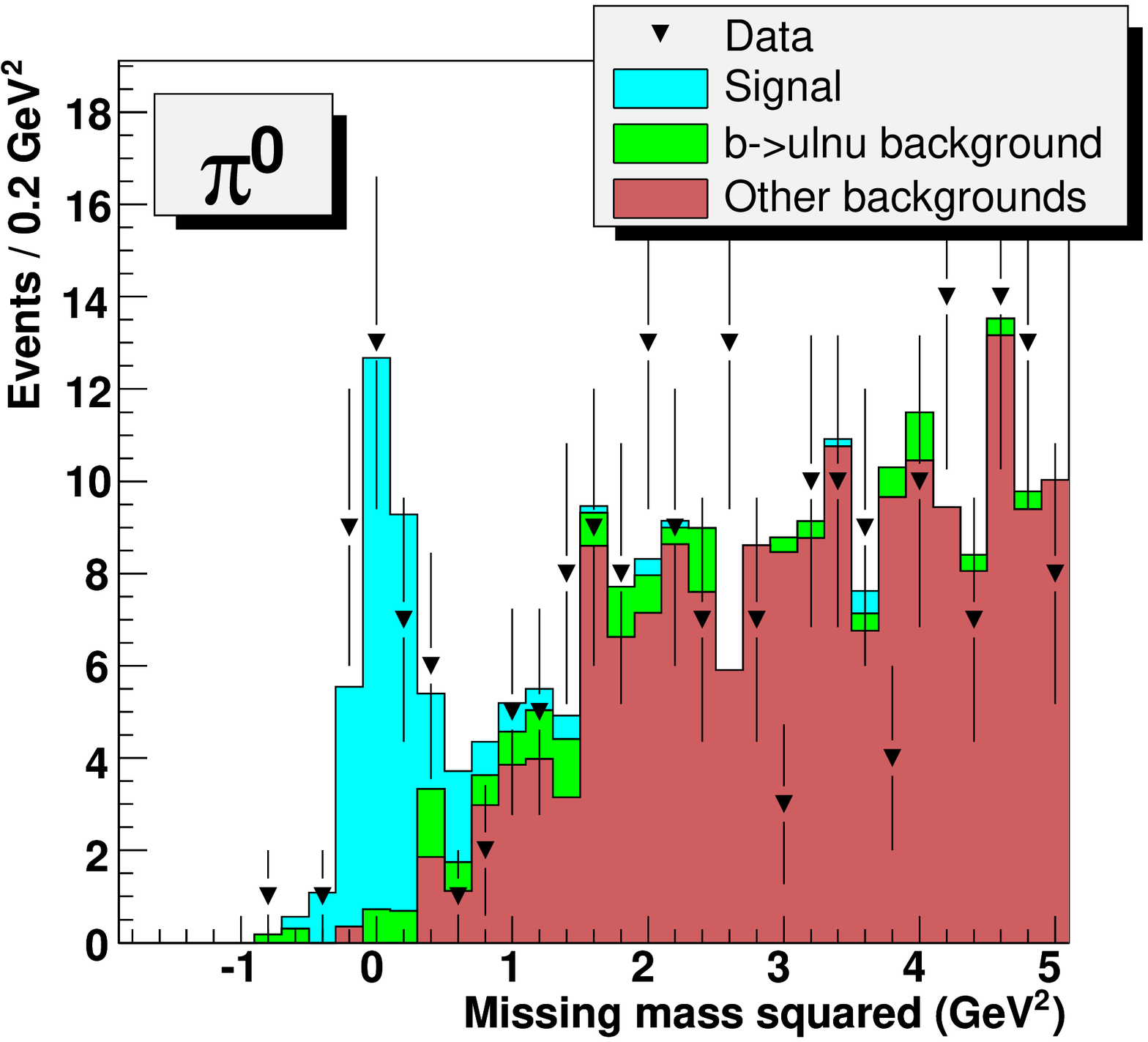} \\
  (c) & (d) \\
\includegraphics[width=0.35\textwidth]{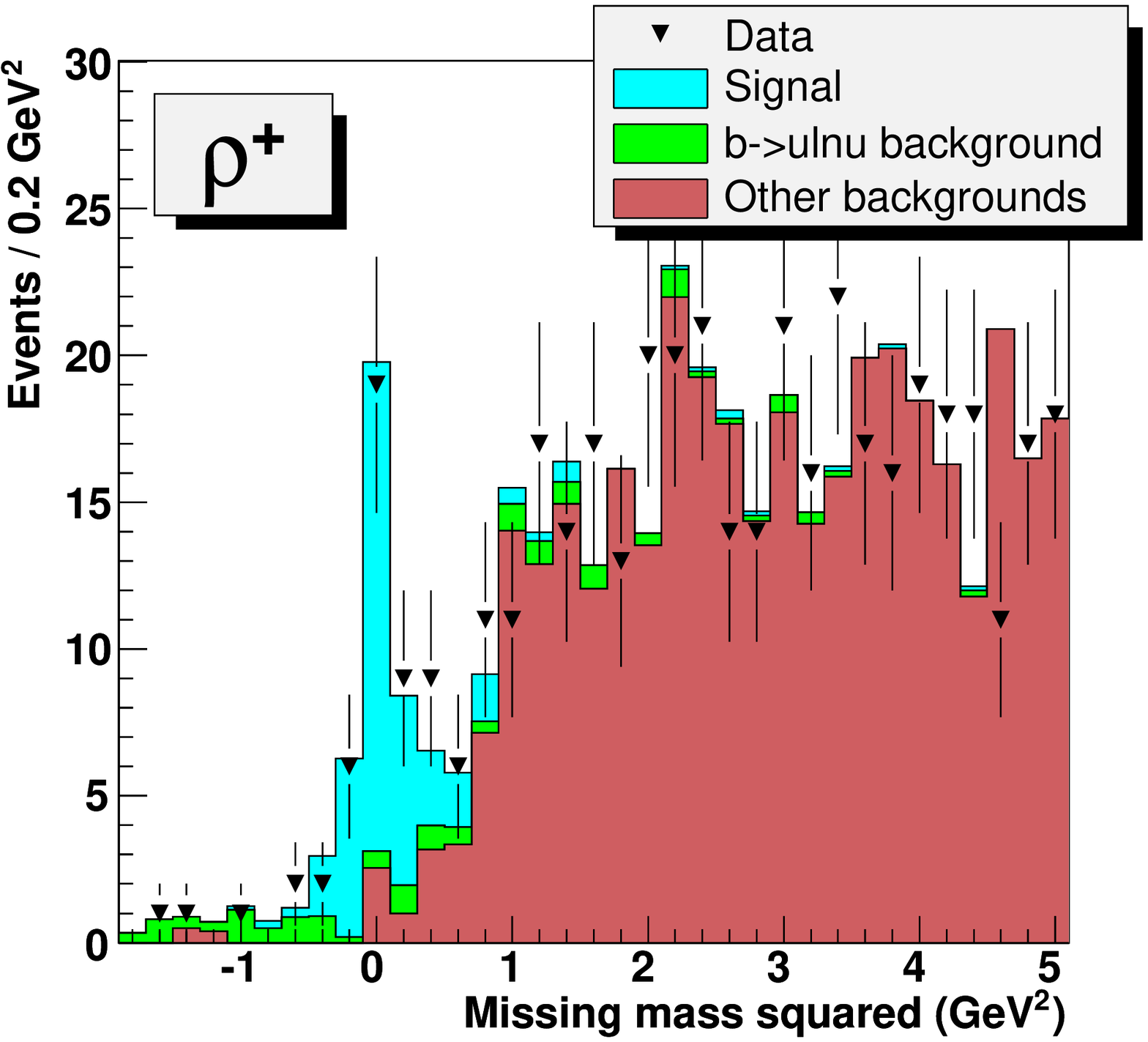} &
\includegraphics[width=0.35\textwidth]{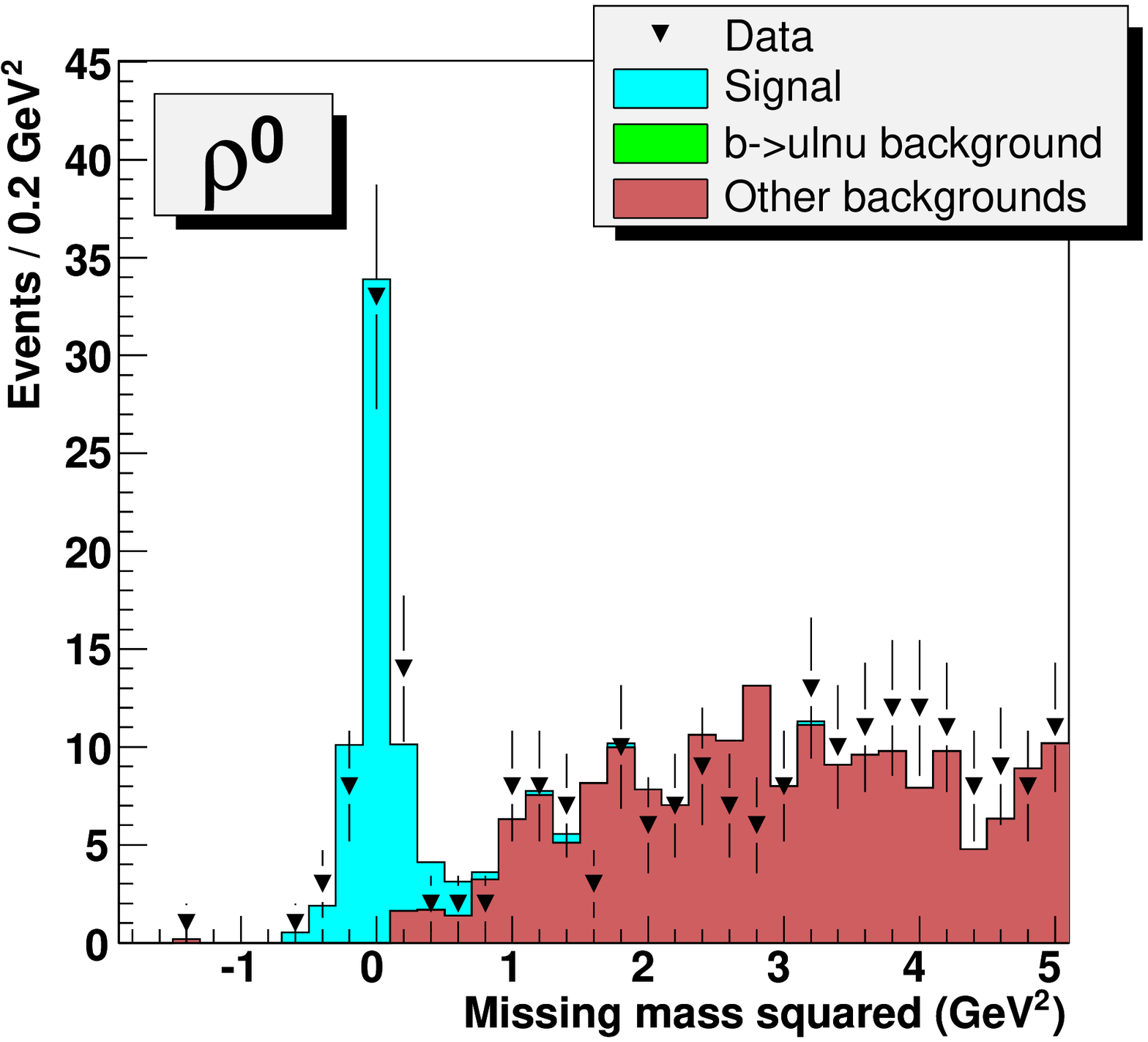} \\
  (e) &     \\
\includegraphics[width=0.35\textwidth]{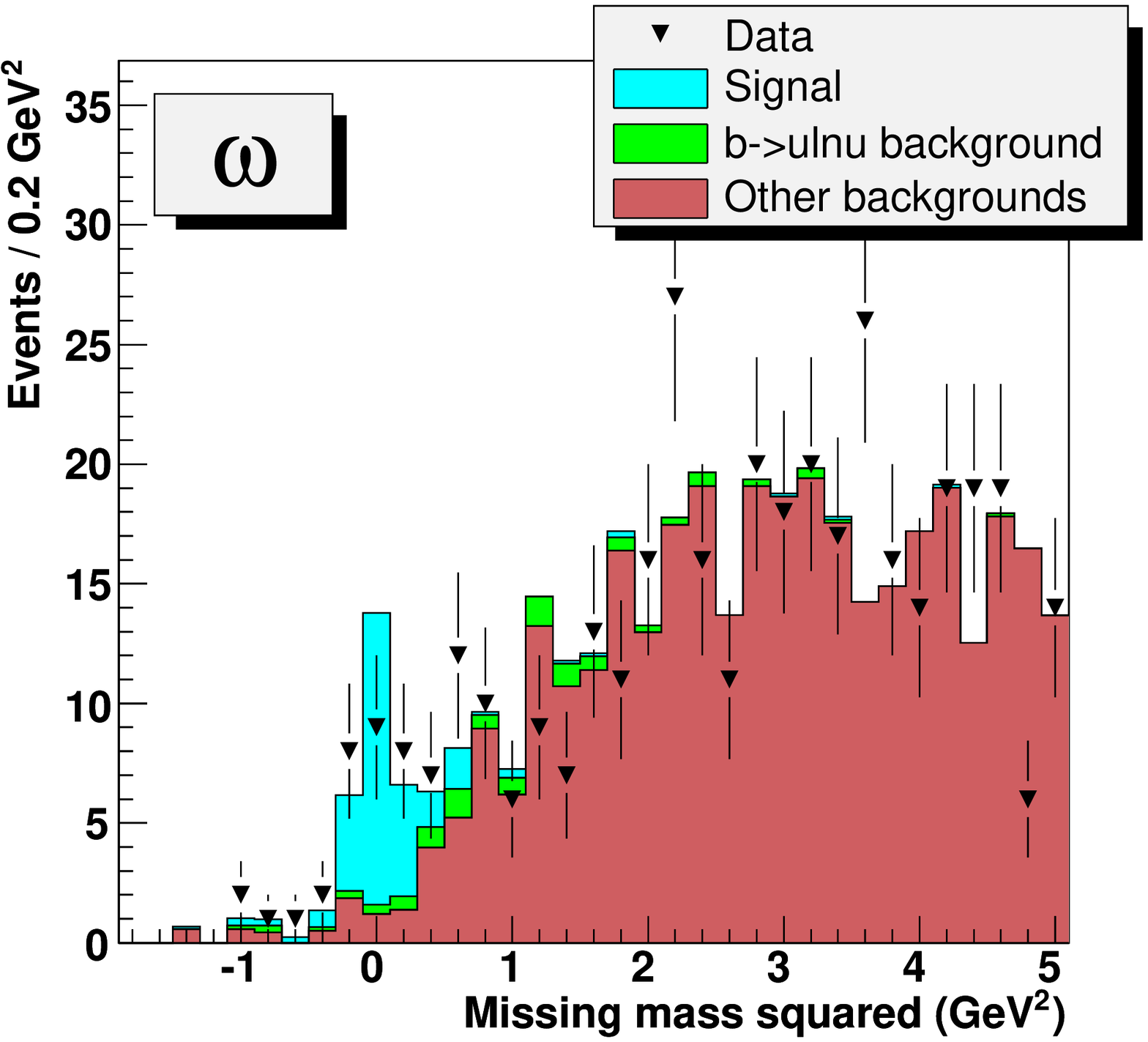} &
\end{tabular}
\caption{Missing mass squared (\mm2)
  distributions after
all cuts, for (a) \bpiplnu, (b) \bpizlnu, (c) \brhoplnu, (d) \brhozlnu,
and (e) \bomegalnu\ modes.
Data is indicated by the points with error
bars. The blue histogram (lightest shade in greyscale) shows the
fitted prediction based on the LCSR
model \cite{ball-and-zwicky-01} \cite{ball-and-braun-98}.
The green histogram (middle shade in greyscale) shows the fitted
$b \to u \ell \nu$ background
contribution. The crimson histogram (darkest shade in greyscale) shows
the fitted background contribution from other sources. The fitting
method is explained in the text.}
\end{figure}

Figure~\ref{fig:mm2fit} shows the observed \mm2\ distributions for the
five decay modes. Also shown in this figure are the fitted signal
component, the fitted \ulnu\ crossfeed, and the fitted contribution from other
backgrounds, which is dominated by \bxclnu\ decays. The shapes of
these components are taken from MC and the normalizations are fit
parameters. The fitting method follows that of Barlow and Beeston
\cite{barlow-and-beeston-93} and takes into account finite Monte Carlo
statistics. The fitted event yields obtained are $48 \pm 8$ for the
\bpiplnu\ mode, $35 \pm 7$ for \bpizlnu, $41 \pm 9$ for \brhoplnu,
$61 \pm 9$ for \brhozlnu\ and  $27 \pm 9$ for \bomegalnu.

We extract the partial branching fractions in bins of \qsq, the
invariant mass squared of the lepton-neutrino system, in order to
minimise the systematic error which arises from the lack of precise
knowledge of the shape of the form factors. These are shown in
Figure~\ref{fig:mm2fit}.
Three bins of \qsq\  are chosen, commensurate with available
statistics, $0 \mathrm{\ to \ } 8\ \GeVGeVcc$,
$8 \mathrm{\ to \ } 16\ \GeVGeVcc$, and greaterthan $16\ \GeVGeVcc$.
The neutrino 4-momentum is determined from the missing 4-momentum
vector using
$p_{\nu} = \left( |\vec{p}_\mathrm{miss}|, \vec{p}_\mathrm{miss} \right)$, 
where $\vec{p}_\mathrm{miss}$ is the missing 3-momentum vector in the
$\Upsilon(4S)$ rest frame.
The \qsq\  resolution obtained varies from $0.20\ \GeVGeVcc$ for
the \bpiplnu\ channel to $0.27\ \GeVGeVcc$ for the \bomegalnu\ channel.


Table~\ref{table:systsum}
summarises the result of a preliminary study of the contributions to
the total systematic error for the branching fractions summed over the
three \qsq\ bins, for the \bpiplnu\ and \bpizlnu\ modes.
These are broken down into the following categories;
those arising from detector simulation, such as charged track
reconstruction efficiency, particle identification and neutral cluster
reconstruction; uncertainties in the luminosity; and effects of the
form factor models used and assumed branching fractions in the MC.

The effects of model dependence of the form factor shapes assumed in the
\bxulnu\ MC used for signal efficiency and crossfeed background estimates
have been studied by comparing the fitted yields obtained using the
default model implemented in the MC, which is LCSR
\cite{ball-and-braun-98} \cite{ball-and-zwicky-01}, and the ISGW2
model \cite{isgw2}. This is achieved by reweighting the MC events
on an event-by-event basis based on their generated values of \qsq\
and angular variables.  The variation between these two models in
predicting the shapes of the \qsq\ distributions for the pseudoscalar
and vector modes typifies the spread between available models for the
dynamics of these decays. 

Effects due to the uncertainties in the branching fraction
normalisations for \btoulnu\ and \btoclnu\ decays in the signal
and background MC samples were studied by
varying in turn the \bpiplnu, \bpizlnu, \brhoplnu,
\brhozlnu, \bomegalnu, \bdplnu, \bdzlnu, \bdsplnu\ and \bdszlnu\
branching fractions by
their measurement errors as quoted by the Particle Data
Group~\cite{PDG}. A reweighting technique is again used, and 
fitted yields with and without reweighting are compared. The maximum
observed spread in the fitted branching fraction is assigned as
systematic error.

The effects of finite MC statistics are taken into account in the
fitting procedure \cite{barlow-and-beeston-93} and are reflected in the
errors on the obtained branching fractions. Since the available MC
samples are rather limited in statistics, variations of the
assumptions on form factor shapes and normalizations can be absorbed
by the present fits to a significant extent.

\begin{table}[htb]
\caption{Results of a preliminary study of sources of systematic uncertainty.}
\label{table:systsum}
\begin{tabular}
      {@{\hspace{0.35cm}}l@{\hspace{0.35cm}}  
       @{\hspace{0.35cm}}c@{\hspace{0.35cm}}  
       @{\hspace{0.35cm}}c@{\hspace{0.35cm}}}  
  \hline \hline
 Source of error & \multicolumn{2}{c}{Assigned systematic error} \\
  & \bpiplnu & \bpizlnu  \\
 \hline \hline
     {\bf Detector Simulation:} & &  \\
     Pion track finding eff. & 1.3\% & -   \\
     $\pi^0$ reconstruction eff. & - & 4.6\%  \\
     Lepton track finding eff. & 1\% & 1\%  \\
     Lepton identification & 2.1\% & 2.1\%  \\
     Charged kaon identification & 2.0\% & -  \\
     Combined & 3.3\% & 5.2\%  \\
     \hline
     N(\bb) uncertainty & \multicolumn{2}{c}{1.3\%} \\
     \hline 
	 {\bf Form Factor Shapes:}  & &  \\
	 $\pi$ (LCSR $\to$ ISGW2) & 2.2\% & 1.4\%  \\
	 $\rho,\omega$ (LCSR $\to$ ISGW2)
	                          & 0.1\% & 2.4\%  \\
     \hline 
	 {\bf Branching Fractions:} & &  \\
	 \btoulnu & 0.0\% & 2.2\%  \\
	 \btoclnu & 0.7\% & 2.2\%  \\
  \hline 
 {\bf Total systematic error} &  4.2\% & 6.8\%  \\
  \hline \hline
\end{tabular}
\end{table}

\begin{figure}[h]
  \begin{tabular}{ll}
    (a) & (b) \\
\includegraphics[width=0.35\textwidth]{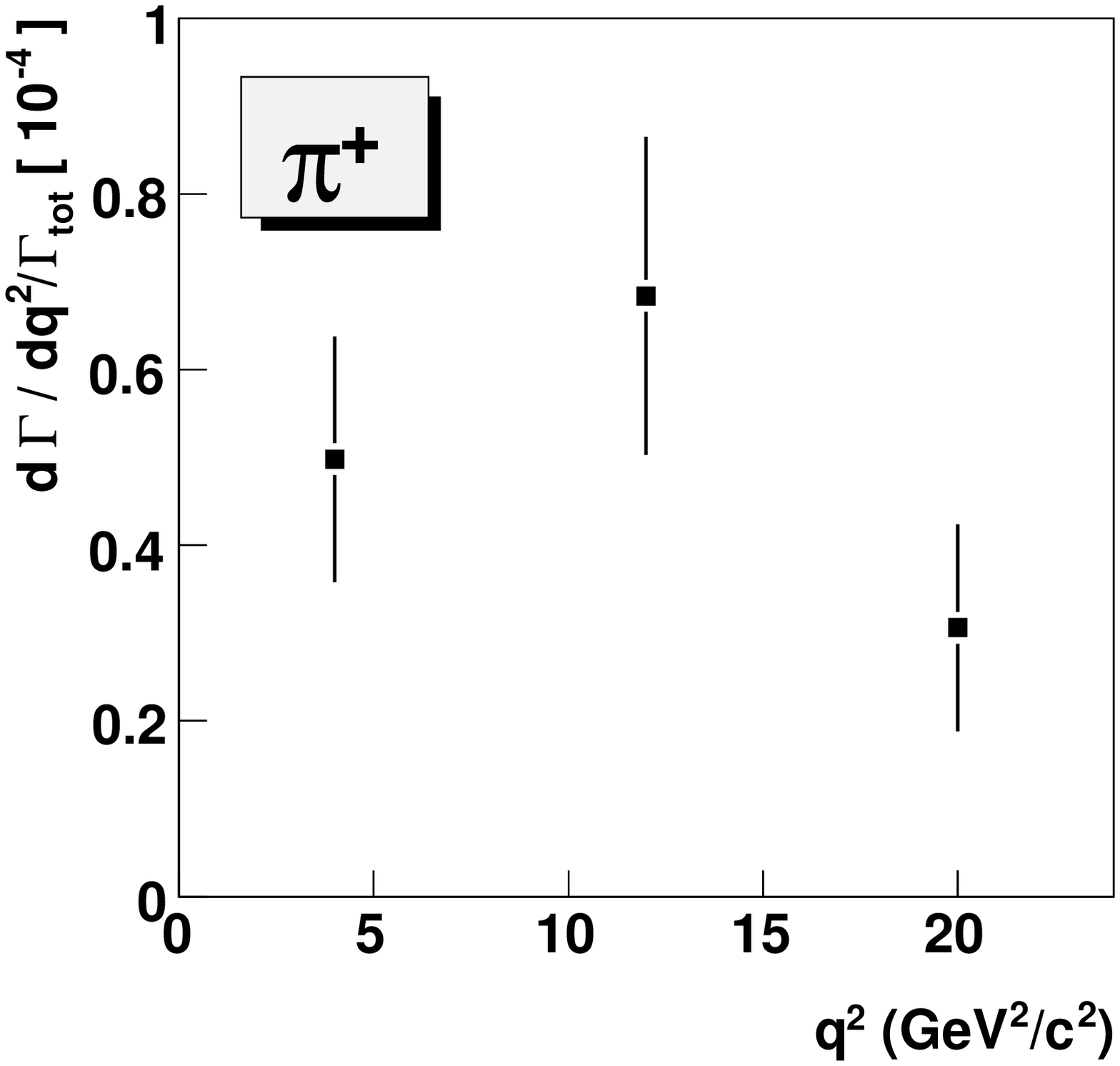}  &
\includegraphics[width=0.35\textwidth]{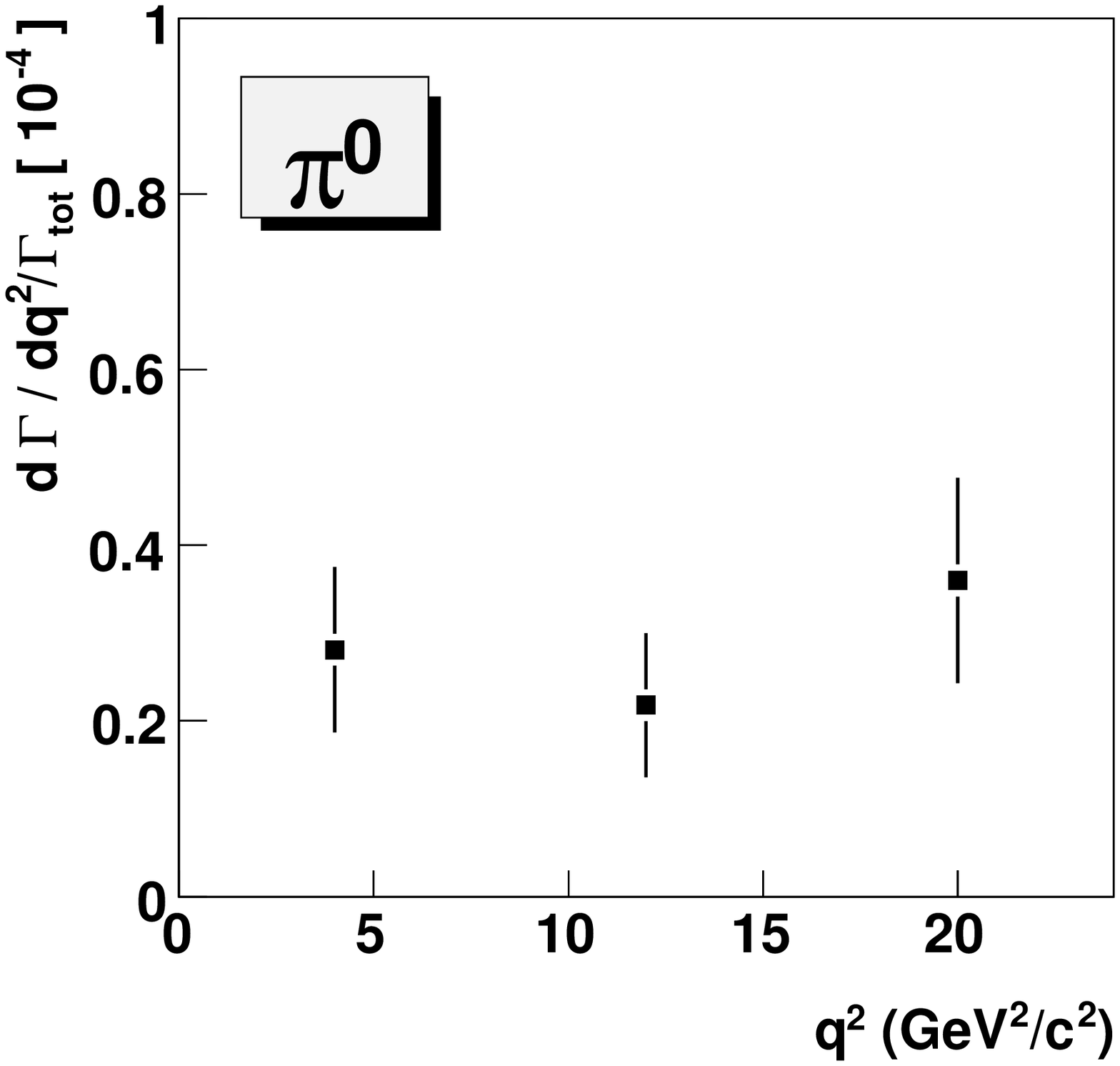}  \\
    (c) & (d) \\
\includegraphics[width=0.35\textwidth]{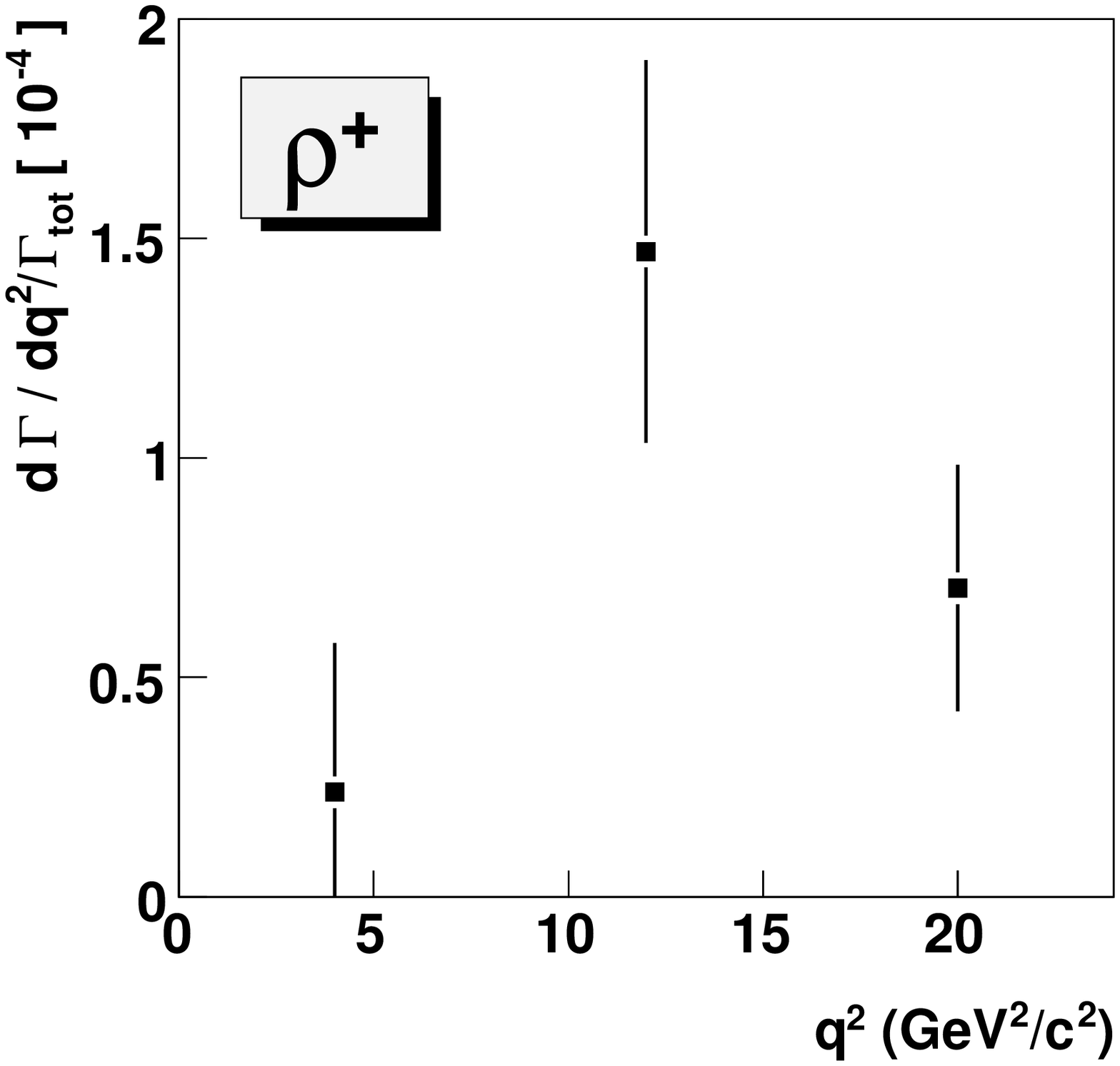}  &
\includegraphics[width=0.35\textwidth]{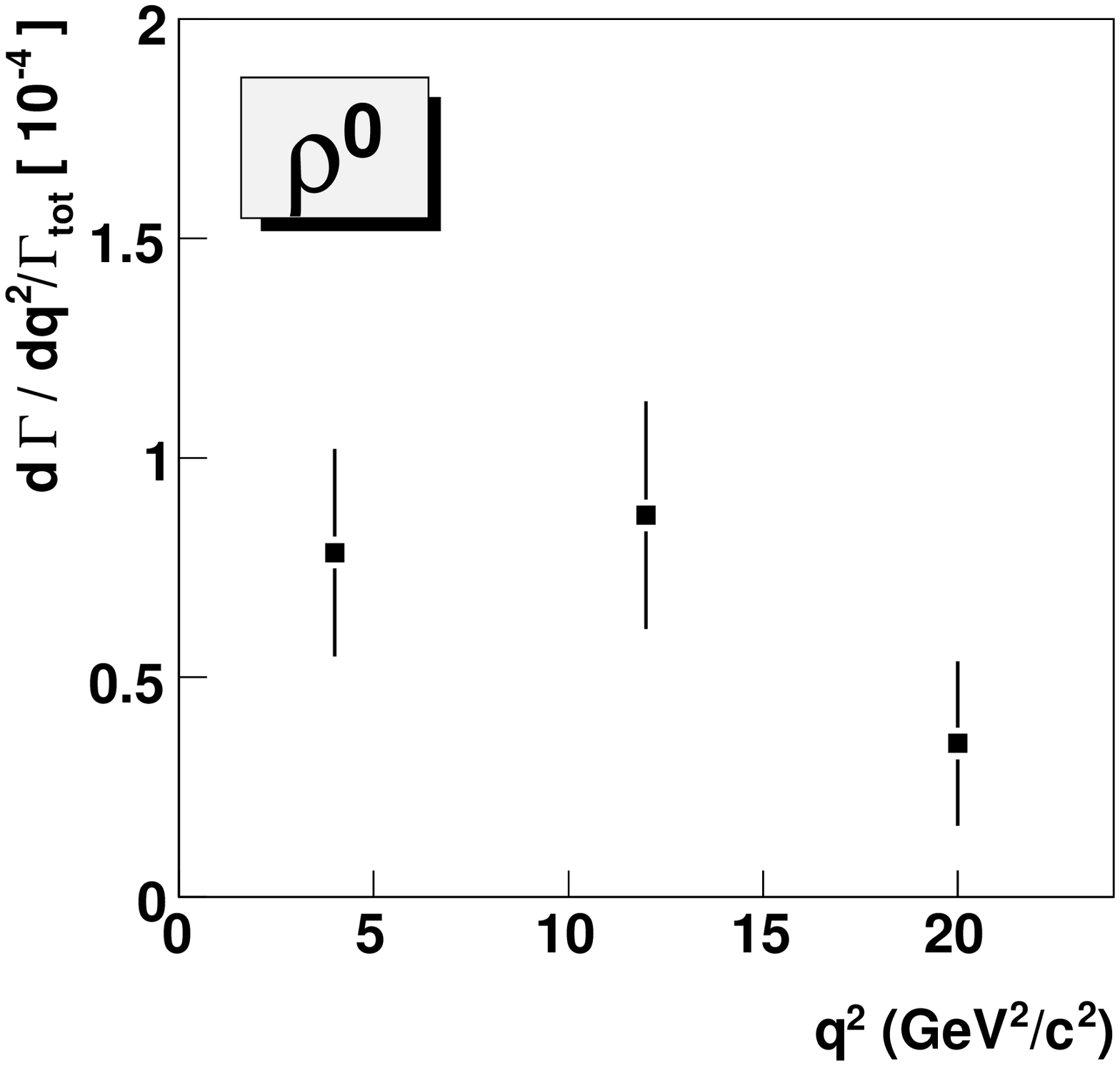}  \\
    (e) &     \\
\includegraphics[width=0.35\textwidth]{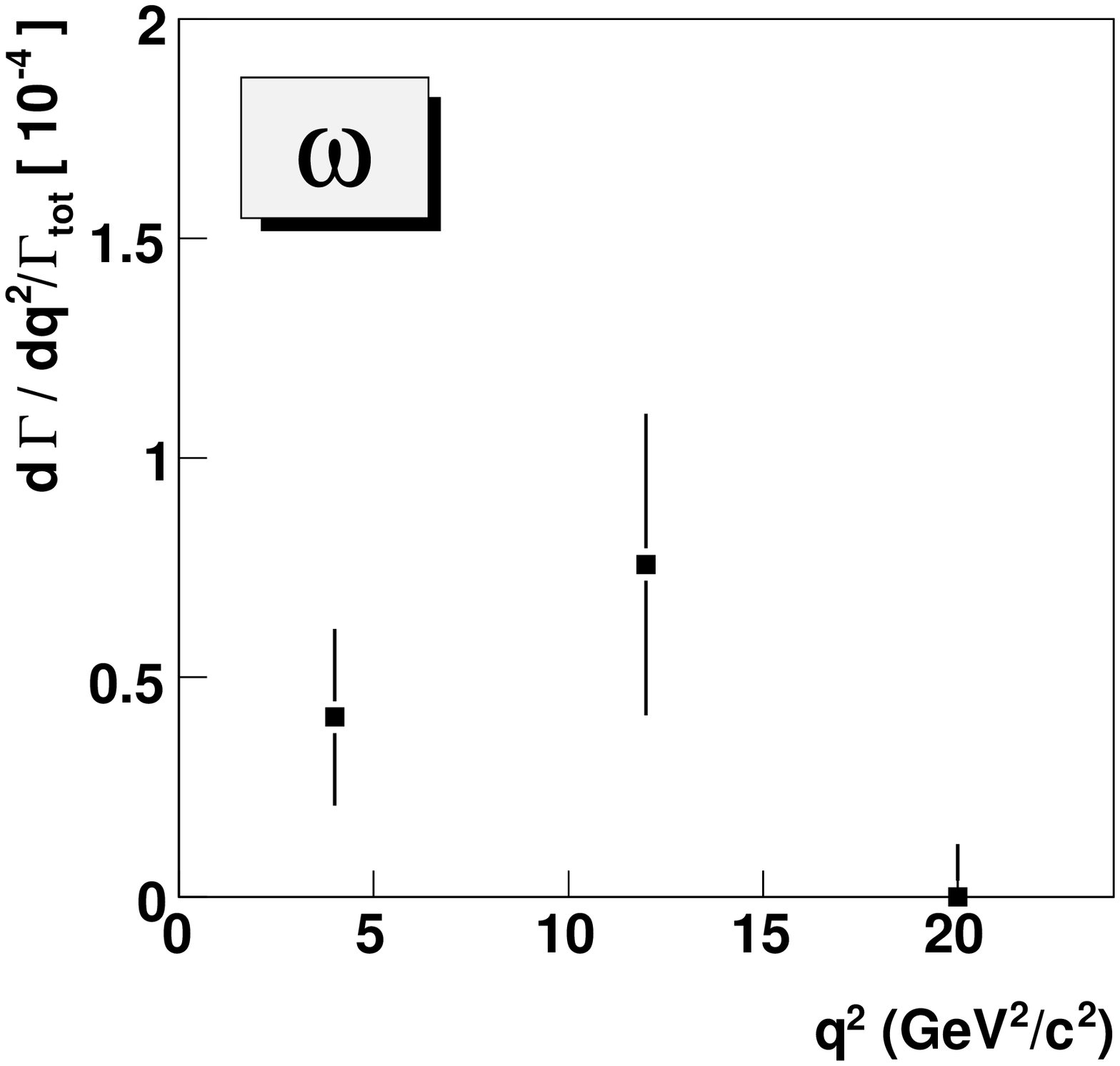}  \\
\end{tabular}
\caption{Partial branching fractions as a function of \qsq\ for the
  five signal modes (a) \bpiplnu, (b) \bpizlnu, (c) \brhoplnu,
  (d) \brhozlnu, and (e) \bomegalnu. Errors shown are statistical and
  preliminary systematic, added in quadrature.}
\label{fig:qsq_bfs}
\end{figure}

The partial branching fractions in bins of \qsq\ are given in
Table~\ref{table:qsq_bfs} for the \bpiplnu\ and \bpizlnu\ modes, for
which reliable preliminary systematic errors have been
estimated. The systematic errors are included for each bin as well
as the sum over bins. 
Figure~\ref{fig:qsq_bfs} presents the shapes of the partial branching
fractions for all five modes as a function of \qsq, where the
statistical and systematic errors have been added in quadrature.
The systematic errors for the vector modes in this figure should
be considered as very preliminary.

\begin{table}[htb]
    \caption{Partial branching fractions in three
      bins of \qsq. These are summed to give the full branching
      fraction quoted in the ``Sum'' column. Errors are statistical
      and systematic.}
    \label{table:qsq_bfs} 
    \begin{tabular}
      {@{\hspace{0.3cm}}c@{\hspace{0.3cm}}  
       @{\hspace{0.3cm}}c@{\hspace{0.3cm}}  
       @{\hspace{0.3cm}}c@{\hspace{0.3cm}}  
       @{\hspace{0.3cm}}c@{\hspace{0.3cm}}  
       @{\hspace{0.3cm}}c@{\hspace{0.3cm}}}
      \hline \hline
      &
      \multicolumn{3}{c}{$\Delta\mathcal{B} \left[ 10^{-4}\right]$} &
      {$\mathcal{B} \left[ 10^{-4}\right]$} \\
      \cline{2-5}
      Mode
      & $0 < q^2 < 8$
      & $8 < q^2 < 16$
      & $q^2 > 16$
      & Sum \\
      & (\GeVGeVcc)
      & (\GeVGeVcc)
      & (\GeVGeVcc)
      & (\GeVGeVcc) \\
      \hline
      \bpiplnu
      & \pipshortresultqsqafit 
      & \pipshortresultqsqbfit
      & \pipshortresultqsqcfit
      & \pipshortresultqsqsumfit \\
      \bpizlnu
      & \pizshortresultqsqafit
      & \pizshortresultqsqbfit
      & \pizshortresultqsqcfit
      & \pizshortresultqsqsumfit \\ 
     \hline
    \end{tabular}
\end{table}

In summary, we have made a preliminary study of the partial branching
fractions as a function of \qsq\ for five semileptonic decay channels
of $B$ mesons to charmless final states, using a full reconstruction
tag method. Summed over the three \qsq\ bins we obtain the following
estimates of the branching fractions for the pion modes: \
$\mathcal{B}\left( \bpiplnu \right)
  = \left( \pipshortresultqsqsumfit \right) \times 10^{-4}$,\
  $\mathcal{B}\left( \bpizlnu \right)
  = \left( \pizshortresultqsqsumfit \right) \times 10^{-4}$,
  where the first error is statistical and the second a preliminary
  estimate of the systematic error. Whilst the statistical precision
  of these measurements is limited at present, the potential power of the
  full reconstruction tagging method, when it can be used with larger
  accumulated $B$-factory data samples in the future, can clearly be seen. 


  \vspace*{\baselineskip}
  
  


\vspace*{2\baselineskip}


We thank the KEKB group for the excellent operation of the
accelerator, the KEK cryogenics group for the efficient
operation of the solenoid, and the KEK computer group and
the National Institute of Informatics for valuable computing
and Super-SINET network support. We acknowledge support from
the Ministry of Education, Culture, Sports, Science, and
Technology of Japan and the Japan Society for the Promotion
of Science; the Australian Research Council and the
Australian Department of Education, Science and Training;
the National Science Foundation of China and the Knowledge
Innovation Program of the Chinese Academy of Sciencies under
contract No.~10575109 and IHEP-U-503; the Department of
Science and Technology of India; 
the BK21 program of the Ministry of Education of Korea, 
the CHEP SRC program and Basic Research program 
(grant No.~R01-2005-000-10089-0) of the Korea Science and
Engineering Foundation, and the Pure Basic Research Group 
program of the Korea Research Foundation; 
the Polish State Committee for Scientific Research; 
the Ministry of Science and Technology of the Russian
Federation; the Slovenian Research Agency;  the Swiss
National Science Foundation; the National Science Council
and the Ministry of Education of Taiwan; and the U.S.\
Department of Energy.


%


\begin{thebibliography}{99}
\bibitem{CC}
Throughout this paper, 
the inclusion of the charge conjugate mode decay is implied
unless otherwise stated.
\end{thebibliography}
\end{document}